# Breaking the Sabatier Principle by Dynamic Adsorption–Desorption Decoupling in Electrocatalytic Hydrogen Evolution

Zi-Xuan Yang[1], Lei Li[1], Tao Huang[2,1], Hui Wan[3,1], X. S. Wang[1*], Gui-Fang Huang[1*], Wangyu Hu[4], Wei-Qing Huang[1*]

[1] Department of Applied Physics, School of Physics and Electronics, Hunan University, Changsha 410082, China
[2] School of Materials Engineering, Xihang University, Xian, 710077, China.
[3] School of Materials and Environmental Engineering, Changsha University, Changsha, 410082, China.
[4] College of Materials Science and Engineering, Hunan University, Changsha, 410082, China.

**Abstract**: The Sabatier principle establishes a fundamental trade-off in heterogeneous electrocatalysis. In the hydrogen evolution reaction (HER), this trade-off is manifested by the coupling of Volmer step, which requires strong hydrogen adsorption, with the Heyrovsky/Tafel step, which favors facile desorption—thus giving rise to the classical volcano relationship and limiting activity even at $\Delta G_{H^*} \approx 0$. Here, we demonstrate a ferroelectric platform with dynamic tunability—monolayer $GeS_2$ decorated with transition metal atoms as a proof-of-concept — where polarization-driven surface electronic reconstruction enables real-time modulation of intermediate binding strength, thereby breaking the Sabatier constraint. Reversible control of hydrogen adsorption allows strong H* binding to accelerate the Volmer step, followed by weakened adsorption to promote the Heyrovsky/Tafel step. This dynamic adsorption–desorption decoupling not only surpasses the volcano limit to achieve unprecedented HER activity, but also establishes a general paradigm for designing adaptive electrocatalysts capable of reconfiguring under operating conditions.

[*]. *E-mail addresses:* wqhuang@hnu.edu.cn, justicewxs@hnu.edu.cn, gfhuang@hnu.edu.cn

**Introduction**

The Sabatier principle has long been regarded as a cornerstone in the rational design of heterogeneous electrocatalysts.[1–3] It postulates that an ideal catalyst should bind reaction intermediates with moderate strength—neither too strongly, which hinders product desorption, nor too weakly, which impedes intermediate adsorption.[4–7] This intrinsic trade-off results in the well-known volcano-type activity relationship, which has been widely observed in various electrochemical reactions,[8–10] including the hydrogen evolution reaction (HER).[11–13] In HER, the initial Volmer step ($H^+ + e^- + * \rightarrow H*$) requires sufficiently strong hydrogen adsorption to facilitate proton–electron transfer, whereas the subsequent Heyrovsky ($H* + H^+ + e^- \rightarrow H_2 + *$) or Tafel ($2H* \rightarrow H_2 + 2*$) step demands facile hydrogen desorption.[14–16] The coupled nature of these steps creates a fundamental kinetic bottleneck that limits further performance improvement, even for catalysts with near-optimal hydrogen adsorption free energy ($\Delta G_{H^*} \approx 0$)[17–19].

The HER is a prototypical two-electron transfer reaction involving a single key intermediate, $H*$, and proceeds via either the Volmer–Heyrovsky or Volmer–Tafel pathway.[14,15,20] When $\Delta G_{H^*}$ deviates significantly from zero, either the adsorption-limited Volmer step or the desorption-limited Heyrovsky/Tafel step dominates, leading to suboptimal kinetics.[21,22] As shown in Fig. 1a, this relationship is quantitatively captured by volcano plots, where experimentally measured exchange current densities correlate with calculated $\Delta G_{H^*}$ values.[23] While this descriptor-based framework effectively identifies the binding characteristics of optimal HER catalysts, it cannot circumvent the inherent coupling between adsorption and desorption steps (Fig. 1b). Moreover, additional factors—such as kinetic barriers[24–26], surface state changes[27–29], and pH-dependent proton transfer dynamics[30–32]—further complicate the optimization process and underscore the limitations of the Sabatier principle in guiding the design of next-generation HER catalysts.

To transcend this intrinsic constraint, a paradigm shift is required—one that enables independent modulation of adsorption and desorption energetics. Here, we report a proof-of-concept catalyst system in which in-plane ferroelectric polarization drives surface electronic reconstruction, thereby providing real-time control over intermediate binding and dynamically decoupling the Volmer and Heyrovsky/Tafel steps. Polarization reversal modulates the orbital hybridization between the $GeS_2$ support and anchored transition-metal (TM) single atoms, reconfiguring the localization and energy distribution of their *d* orbitals. First-principles calculations demonstrate that this strategy tunes the hydrogen adsorption free energy from a nearly fixed value (~0 eV) typical of conventional catalysts to a polarization-decoupled adsorption–desorption process. The optimal system, W@GeS$_2$ enables the $\Delta G_{H^*}$ to swing between −0.6 and 0.9 eV, thereby establishing a "strong–weak" alternating adsorption mechanism. This dynamic adsorption–desorption modulation not only circumvents the Sabatier limitation and surpasses the volcano limit, but also offers a generalizable paradigm for designing next-generation electrocatalysts capable of adaptive reconfiguration under operating conditions.

**Result and Discussion**

To dynamically decouple hydrogen adsorption and desorption during the HER, we harness ferroelectric polarization as a reversible strategy to modulate the binding energetics of active center in real time (Fig. 1c).[33–35] Under the FE1 polarization state, polarization-induced electronic reconstruction strengthens the H*–active center interaction, lowering the adsorption free energy and promoting efficient proton–electron transfer in the Volmer step (Fig. 1d, left). Reversing the polarization to the FE2 state weakens the binding, shifting the adsorption free energy toward positive values and facilitating hydrogen release through the Tafel or Heyrovsky step (Fig. 1d, right). Such polarization-controlled switching between strong and weak adsorption states establishes a dynamic "strong–weak" modulation, effectively circumventing the Sabatier limitation and enabling real-time optimization of successive HER steps.

To establish the polarization-regulated HER pathway, we design a proof-of-concept system by anchoring single transition-metal atoms (TM = Ti–Cu) on a ferroelectric GeS$_2$ substrate (Fig. 2a). GeS$_2$ serves as an ideal ferroelectric support due to its robust in-plane polarization (107 μC/cm², Fig. S1), low switching barrier (399 meV), and remarkable thermal stability up to 1000 K (Fig. S2). The polarization reversal originates from the displacement of Ge atoms along the [110] direction (Fig. 2a).[36]

In the TM@GeS$_2$ system, the switching of in-plane ferroelectric polarization propagates into the vertical direction, leading to a pronounced modulation of the out-of-plane dipole moment ($\mu$). The magnitude of $\mu$ depends on the charge variation ($q$) of the TM atom and the vertical thickness ($d$) of the system, which can be approximated as $\mu \sim q \cdot d$.[37] Structural analysis reveal three key effects during the in-plane polarization switching (Fig. 2b): (i) Pronounced out-of-plane lattice strain: upon switching from the FE1 to the FE2 state, $d$ increases, with the most notable change observed in Cr@GeS$_2$ (from 3.78 Å to 5.26 Å). We attribute this strain to the strong electrostatic repulsion due to the displacement of Ge atom, which align more directly with the TM atom. (ii) Charge redistribution: In Co- and Ni@GeS$_2$, the charge transfer associated with the TM atom is markedly enhanced under the FE2 state relative to FE1, whereas the opposite trend is observed in other systems. This highlights a strong correlation between polarization switching and the electronic states of the TM atom. (iii) Enhancement of $\mu$: $\mu$ is generally larger under the FE2 state compared to FE1. The most prominent case is Ti@GeS$_2$, where $\mu$ increases from -0.53 e·Å (FE1) to -1.26 e·Å (FE2) As a result, these structural and electronic evolutions demonstrate that in-plane polarization switching not only induces substantial structural deformation but also drives a reconstruction of the out-of-plane electronic structure. This is manifested in the enhancement of the dipole moment and the redistribution of the TM-atom charge, which collectively modulate the hydrogen adsorption energy.

We systematically evaluate the HER performance of eight TM@GeS$_2$ configurations under two polarization states (FE1 and FE2) (Fig. 2c and Table S1), and found that hydrogen adsorption is highly sensitive to the polarization state. In the Ti-, V-, Cr-, and Mn@GeS$_2$ systems, the FE1 state exhibits a lower $\Delta G_{H^*}$. Taking V@GeS$_2$ as an illustrative example, $\Delta G_{H^*}$ is of –0.01 eV under the FE1 state, favoring hydrogen adsorption (Volmer step), whereas it increases substantially to 0.60 eV under the FE2 state, which facilitates subsequent desorption (Heyrovsky/Tafel step). This polarization-dependent switching thus enables a decoupling of the adsorption and desorption processes. In contrast, the Fe-, Co-, Ni-, and Cu@GeS$_2$ systems exhibit smaller $\Delta G_{H^*}$ values under the FE2 state than under the FE1 state, but neither polarization state provides a sufficiently favorable adsorption energy.

To unravel the physical origin of this polarization-dependent trend, we first examine the correlation between $\Delta G_{H^*}$ and two dipole-related descriptors: the out-of-plane dipole moment ($\mu$), the local dipole moment ($\mu_d$) arising from charge transfer between the TM atom and the GeS$_2$ (Fig. S3). However, neither $\mu$ nor $\mu_d$ accounts for the observed variation in $\Delta G_{H^*}$. As shown in Fig. S3a2 and b2, increasing $\Delta\mu$ or $\Delta\mu_d$ does not coincide with an increase in $\Delta G_{H^*}$. To gain deeper insight, we turn to an orbital interaction analysis. As depicted in Fig. 2d, polarization switching modulates the coupling between the TM atom and the GeS$_2$ host, thereby reshaping the electronic configuration of the TM $d_{z^2}$ orbital. This orbital reorganization in turn alters the σ-type overlap between the H 1$s$ and TM $d_{z^2}$ orbitals, enabling a controllable transition from strong to weak coupling states. Such orbital-level modulation provides a more comprehensive physical picture for the polarization control of hydrogen adsorption energetics.

Interestingly, we observe a discontinuity in the projected density of states (PDOS) near the Fermi level for the TM atoms in the Ti-, V-, Cr-, Mn-, and Co@GeS$_2$ systems under specific ferroelectric polarization state, predominantly governed by the TM $d_{z^2}$ orbital. As shown in Fig. S4, such band splitting occurs under the FE2 state for Ti-, V-, Cr-, and Mn@GeS$_2$ systems, whereas Co@GeS$_2$ exhibits this feature under the FE1 state. This polarization-induced electronic reconstruction originates from modifications in the local coordination environment. As illustrated in Fig. 3a, the FE1→FE2 switching drives Ge atom to align directly beneath the TM atom, thereby elongating the TM–S bond and reducing the S–TM–S bond angle. This structural reconfiguration enhances the hybridization among the TM-$d_{z^2}$, S-$p_z$, and Ge-$d_{z^2}$ orbitals under the FE2 state, ultimately leading to energy level splitting of the TM $d_{z^2}$ orbital, with the magnitude of this splitting denoted as $\Delta_{o(d_{z^2})}$. Notably, in the Ti@GeS$_2$ (Fig. 3b), ferroelectric polarization switching also modulates the spin polarization of the TM atom. Under the FE1 state, the $d_{z^2}$ orbital is spin-polarized, predominantly occupied by spin-up electrons; however, this spin polarization vanishes under the FE2 state. Such spin-state transitions further regulate the interaction between H and the TM site.

Building on the reconstructed $d_{z^2}$ orbital, we find that configurations with pronounced orbital splitting generally exhibit weaker hydrogen adsorption, while continuous $d_{z^2}$ states facilitate more efficient charge transfer and stronger H binding. This mechanism rationalizes why Ti-, V-, Cr-, and Mn@GeS$_2$ display weaker adsorption under the FE2 state, whereas Co@GeS$_2$ exhibits reduced binding strength under the FE1 state (Fig. 2c). Further quantitative analysis (Fig. 3c) reveals that among these five systems, Cr@GeS$_2$ has the smallest orbital splitting ($\Delta_{o(d_{z^2})}$ = 0.28 eV), corresponding to a minimal free-energy difference between polarization states (Fig. 3d). By contrast,

V@GeS$_2$ shows the largest splitting ($\Delta_{o(d_{z^2})}$ = 0.84 eV), correlating with the largest free-energy change (~ 0.6 eV) (Fig. 3d). These results establish a direct link between the $\Delta_{o(d_{z^2})}$ and the polarization-switching-induced free-energy difference. A larger $\Delta_{o(d_{z^2})}$ shifts the unoccupied TM-$d_{z^2}$ orbital upward in energy, suppressing hybridization with the H-1$s$ orbital and thereby weakening electron transfer from H to the TM atom. Consequently, the H–TM interaction is significantly reduced.

In TM@GeS$_2$ systems where the TM atoms have more than half-filled $d$ orbitals, the aforementioned orbital-splitting mechanism is no longer valid. As shown in Fig. S5, in Fe-, Co-, Ni-, and Cu@GeS$_2$, the spin-up $d_{z^2}$ orbital is fully occupied, and electrons progressively fill the spin-down channel. In this case, the variation of hydrogen adsorption energy with polarization switching is primarily governed by the energy shift of the TM-$d_{z^2}$ orbital. Consequently, the $d$-band center ($\varepsilon$, calculated according to Eq. (1)) is conventionally employed as a key electronic descriptor of adsorption behavior.

$$\varepsilon = \frac{\int_{-\infty}^{+\infty} n(\varepsilon)\varepsilon\, d\varepsilon}{\int_{-\infty}^{+\infty} n(\varepsilon)\, d\varepsilon} \quad (1)$$

When the H-1$s$ orbital interacts with the TM-$d_{z^2}$ orbital (Fig. 4a1), hybridization occurs, giving rise to a low-energy bonding σ state and a high-energy antibonding σ* state. The occupation of the σ* state directly regulates the stability of the TM–H bond: when the σ* state shifts further away from the Fermi level, its occupation increases, thereby weakening the TM–H interaction. However, in these four systems, the shift of the $d_{z^2}$-band center ($\Delta\varepsilon = \varepsilon(\text{FE2}) - \varepsilon(\text{FE1})$) does not show a clear proportionality with the adsorption free-energy change $\Delta G$ ($G(\text{FE2}) - G(\text{FE1})$) across different polarization states (Fig. S6a). Spin-resolved analysis of the $d_{z^2}$ orbitals (Fig. S6b) reveals that the spin-down channel follows the same trend as $\Delta G$, i.e., a larger $\Delta\varepsilon$ corresponds to a greater $\Delta G$. In contrast, the spin-up channels of Co- and Fe@GeS$_2$ (pink data points) significantly deviate from this trend. This indicates that the $d_{z^2}$-band center alone cannot fully capture the polarization-induced electronic reconstruction.

It is important to note that polarization switching not only shifts orbital energies but also alters orbital occupancies, and both effects jointly dictate hydrogen adsorption. To comprehensively capture this coupling, we introduce the orbital occupancy ($\eta$) as an additional descriptor (Eq. (2)):

$$\eta = \frac{\int_{-\infty}^{0} n(\varepsilon)\, d\varepsilon}{\int_{-\infty}^{+\infty} n(\varepsilon)\, d\varepsilon} \qquad (2)$$

As illustrated in Fig. 4a2, an increased $\eta$ enhances σ* occupation, thereby strengthening electrostatic repulsion between the adsorbate and the TM atom, which weakens the TM–H bond. Accordingly, a larger polarization-induced change in occupancy ($\Delta\eta = \eta(\text{FE2}) - \eta(\text{FE1})$) leads to a greater $\Delta G$.

On this basis, we define a coupled descriptor, $\Delta\varphi = \Delta\varepsilon \times \Delta\eta$, which demonstrates strong correlation with $\Delta G$. As shown in Fig. 4b, among the four systems, Co@GeS$_2$ exhibits the smallest $\Delta\varphi$ (~ –0.50), corresponding to the minimal free-energy change, while Ni@GeS$_2$ shows the largest $\Delta\varphi$ and the largest free-energy change (–0.43 eV). This behavior mainly arises from the spin-down channel (Fig. 4c), which lies closer to the Fermi level and is therefore more sensitive to polarization switching. Therefore, $\Delta\varphi$ effectively captures both the orbital-energy shift ($\Delta\varepsilon$) and occupancy change ($\Delta\eta$), providing a robust descriptor for understanding ferroelectric modulation of hydrogen adsorption energy.

The above analysis primarily addresses the adsorption and desorption of a single hydrogen atom. However, the HER is a catalytic process involving multiple consecutive steps, typically including the Volmer step (adsorption) followed by either the Heyrovsky or Tafel step (desorption). To provide a comprehensive assessment of catalytic performance, we examine the complete HER pathway under both ferroelectric polarization states. As illustrated in Fig. S7, the reaction begins with the adsorption of the first H$^+$ ion on a TM site (Volmer 1). Subsequently, two competing pathways emerge: (i) selective adsorption of a second H$^+$ ion (Volmer 2), or (ii) hydrogen desorption via the Heyrovsky step, combining with another H$^+$ from the solution to form H$_2$. To this end, we calculate the reaction energies for each step across eight systems (Table S2), with energetically favorable pathways highlighted in red. If the system proceeds through Volmer 2, two H atoms desorb simultaneously via the Tafel step, thereby completing the catalytic cycle.

The results reveal that different steps exhibit distinct energy preferences under different polarization states. For instance, in the Fe@GeS$_2$ system, both the Volmer1 and Volmer2 steps are more favorable in the FE2 state, with free energies of –0.14 eV and –0.46 eV, respectively, suggesting enhanced H adsorption under this polarization state. In contrast, the Tafel step is more favorable under the FE1 state (–0.11 eV). These findings demonstrate that ferroelectric switching can dynamically regulate the reaction pathway, allowing it to adapt to the energetically most favorable sequence. As shown in Fig. 5a, applying polarization switching after the Volmer2 and Tafel steps enables a continuous and efficient adsorption–desorption process. Notably, all steps in this pathway exhibit negative free energies, confirming its spontaneous feasibility.

Overall, most 3d TM@GeS$_2$ systems exhibit intrinsically weak hydrogen adsorption. Among them, only V@GeS$_2$ shows polarization-tunable adsorption/desorption behavior, but even under the FE1 state, the H adsorption strength remains suboptimal (-0.01 eV). To identify higher-performance candidates, we extend the search to 4d and 5d periods (Figs. 5b1 and b2). Eight TM@GeS$_2$ systems (TM = Zr, Nb, Mo, Rh, Ta, W, Ir, Au) are identified as promising catalysts with favorable "strong adsorption/weak desorption" characteristics. Seven of these systems require FE1→FE2 switching to complete the catalytic cycle — achieving efficient H adsorption in the FE1 state and facile H$_2$ desorption in the FE2 state; whereas Au@GeS$_2$ follows the opposite FE2→FE1 switching strategy. Remarkably, these high-performance systems exhibit clear periodic trends: Group V elements (V, Nb, Ta) and several Group VI and IX elements (Mo, W, Rh, Ir) all successfully decouple adsorption and desorption, suggesting a strong correlation between HER activity and the group number of the transition metal.

To identify the dominant factors governing $\Delta G$ in these systems, we select 16 candidate descriptors based on structural, electronic, and orbital properties (see definitions and data in Tables S2–S6) and compute their Pearson correlation coefficients with $\Delta G$ (Figs. S8, S9). These descriptors fall into two categories: (i) structure- and charge-based parameters associated with out-of-plane polarization, such as changes in Bader charge variation and vertical thickness; and (ii) spin-polarized orbital parameters, including d-band center shifts and changes in orbital occupancy for different spin channels. Among them, the shift of the spin-down $d_{z^2}$ orbital ($\Delta\varepsilon_{d_{z^2}(D)}$) exhibits the strongest positive correlation with $\Delta G$ (R ≈ 0.42), consistent with the trends observed in Fig. S6, underscoring the dominant role of the spin-down channel in modulating adsorption energetics. In addition, the charge transfer of the TM ($\Delta q_{TM(s)}$) shows a positive correlation with $\Delta G$, highlighting the direct impact of charge redistribution on electrostatic interactions. The thickness along the vertical direction correlates negatively with $\Delta G$ (R ≈ - 0.36), indicating that thicker structures enhance axial polarization effects, promoting electron migration toward active sites and thereby optimizing adsorption. Taken together, these findings establish a comprehensive microscopic framework for understanding ferroelectric-switching-regulated HER activity in TM@GeS$_2$. Specifically, both ferroelectric-structure-driven factors and spin-resolved orbital energetics synergistically dictate hydrogen adsorption free energy, offering predictive insights for the rational design of polarization-tunable electrocatalysts.

**Conclusion**

In summary, we present a ferroelectric polarization-modulation strategy that overcomes the Sabatier constraint in HER by dynamically tuning $\Delta G_{H^*}$. Polarization-driven charge redistribution, structural distortion, and orbital reconfiguration enable reversible control of adsorption strength, reconciling the Volmer and Heyrovsky/Tafel steps. Systematic screening identifies multiple TM@GeS$_2$ systems exhibiting polarization-tunable pathways, demonstrating that ferroelectric catalysts can surpass the static volcano limit and establishing a general paradigm for adaptive electrocatalysis.

## SUPPLEMENTARY MATERIAL

See the supplementary material for the structural details and additional results.

## Acknowledgements

The authors are grateful to the National Natural Science Foundation of China (Grants Nos. 11804045, 12174093 and 52172088), Guangdong Basic and Applied Basic Research Foundation (No. 2024A1515010421), and the Fundamental Research Funds for the Central Universities.

## AUTHOR DECLARATIONS

### Conflict of Interest

The authors have no conflicts to disclose.

## DATA AVAILABILITY

The data that support the findings of this study are available from the corresponding author upon reasonable request.

# References

(1) Shah, A. H.; Zhang, Z.; Huang, Z.; Wang, S.; Zhong, G.; Wan, C.; Alexandrova, A. N.; Huang, Y.; Duan, X. The Role of Alkali Metal Cations and Platinum-Surface Hydroxyl in the Alkaline Hydrogen Evolution Reaction. *Nat. Catal.* **2022**, *5* (10), 923–933.

(2) Hu, S.; Li, W.-X. Sabatier Principle of Metal-Support Interaction for Design of Ultrastable Metal Nanocatalysts. *Science* **2021**, *374* (6573), 1360–1365.

(3) Kari, J.; Olsen, J. P.; Jensen, K.; Badino, S. F.; Krogh, K. B. R. M.; Borch, K.; Westh, P. Sabatier Principle for Interfacial (Heterogeneous) Enzyme Catalysis. *ACS Catal.* **2018**, *8* (12), 11966–11972.

(4) Medford, A. J.; Vojvodic, A.; Hummelshøj, J. S.; Voss, J.; Abild-Pedersen, F.; Studt, F.; Bligaard, T.; Nilsson, A.; Nørskov, J. K. From the Sabatier Principle to a Predictive Theory of Transition-Metal Heterogeneous Catalysis. *J. Catal.* **2015**, *328*, 36–42.

(5) Miao, L.; Jia, W.; Cao, X.; Jiao, L. Computational Chemistry for Water-Splitting Electrocatalysis. *Chem. Soc. Rev.* **2024**, *53* (6), 2771–2807.

(6) Schumann, J.; Stamatakis, M.; Michaelides, A.; Réocreux, R. Ten-Electron Count Rule for the Binding of Adsorbates on Single-Atom Alloy Catalysts. *Nat. Chem.* **2024**, 1–6.

(7) Wang, C.; Wang, X.; Zhang, T.; Qian, P.; Lookman, T.; Su, Y. A Descriptor for the Design of 2D MXene Hydrogen Evolution Reaction Electrocatalysts. *J. Mater. Chem. A* **2022**, *10* (35), 18195–18205.

(8) Huang, T.; Yang, Z.-X.; Li, L.; Wan, H.; Leng, C.; Huang, G.-F.; Hu, W.; Huang, W.-Q. Dipole Effect on Oxygen Evolution Reaction of 2D Janus Single-Atom Catalysts: A Case of Rh Anchored on the P6m2-NP Configurations. *J. Phys. Chem. Lett.* **2024**, *15* (9), 2428–2435.

(9) Wang, Q.; Hung, S.-F.; Lao, K.; Huang, X.; Li, F.; Tao, H. B.; Yang, H. B.; Liu, W.; Wang, W.; Cheng, Y.; Hiraoka, N.; Zhang, L.; Zhang, J.; Liu, Y.; Chen, J.; Xu, Y.; Su, C.; Chen, J. G.; Liu, B. Breaking the Linear Scaling Limit in Multi-Electron-Transfer Electrocatalysis through Intermediate Spillover. *Nat. Catal.* **2025**, *8* (4), 378–388.

(10) Cui, L.; Feng, J.; Wu, S.; Xu, S.; Wu, X.; Zhang, S.; Zhang, Q.; Fu, G.; Wang, Y.; Xie, S. Photocatalyst Design with an Electrical-Polarization-Ability Descriptor for Exclusive C–H Bond Activation in Methanol. *J. Am. Chem. Soc.* **2025**.

(11) Shi, Z.; Zhang, X.; Lin, X.; Liu, G.; Ling, C.; Xi, S.; Chen, B.; Ge, Y.; Tan, C.; Lai, Z.; Huang, Z.; Ruan, X.; Zhai, L.; Li, L.; Li, Z.; Wang, X.; Nam, G.-H.; Liu, J.; He, Q.; Guan, Z.; Wang, J.; Lee, C.-S.; Kucernak, A. R. J.; Zhang, H. Phase-Dependent Growth of Pt on $MoS_2$ for Highly Efficient $H_2$ Evolution. *Nature* **2023**, *621* (7978), 300–305.

(12) Liu, Y.; Ding, J.; Li, F.; Su, X.; Zhang, Q.; Guan, G.; Hu, F.; Zhang, J.; Wang, Q.; Jiang, Y.; Liu, B.; Yang, H. B. Modulating Hydrogen Adsorption via Charge Transfer at the Semiconductor–Metal Heterointerface for Highly Efficient Hydrogen Evolution Catalysis. *Adv. Mater.* **2023**, *35* (1), 2207114.

(13) Er, D.; Ye, H.; Frey, N. C.; Kumar, H.; Lou, J.; Shenoy, V. B. Prediction of Enhanced Catalytic

Activity for Hydrogen Evolution Reaction in Janus Transition Metal Dichalcogenides. *Nano Lett.* **2018**, *18* (6), 3943–3949.

(14) Sun, F.; Tang, Q.; Jiang, D. Theoretical Advances in Understanding and Designing the Active Sites for Hydrogen Evolution Reaction. *ACS Catal.* **2022**, *12* (14), 8404–8433.

(15) Karmakar, A.; Durairaj, M.; Madhu, R.; De, A.; Dhandapani, H. N.; Spencer, M. J. S.; Kundu, S. From Proximity to Energetics: Unveiling the Hidden Compass of Hydrogen Evolution Reaction. *ACS Mater. Lett.* **2024**, *6* (7), 3050–3062.

(16) Kim, H. K.; Jang, H.; Jin, X.; Kim, M. G.; Hwang, S.-J. A Crucial Role of Enhanced Volmer-Tafel Mechanism in Improving the Electrocatalytic Activity *via* Synergetic Optimization of Host, Interlayer, and Surface Features of 2D Nanosheets. *Appl. Catal., B* **2022**, *312*, 121391.

(17) Xu, H.; Cheng, D.; Cao, D.; Zeng, X. C. Revisiting the Universal Principle for the Rational Design of Single-Atom Electrocatalysts. *Nat Catal* **2024**, *7* (2), 207–218.

(18) Boonpalit, K.; Wongnongwa, Y.; Prommin, C.; Nutanong, S.; Namuangruk, S. Data-Driven Discovery of Graphene-Based Dual-Atom Catalysts for Hydrogen Evolution Reaction with Graph Neural Network and DFT Calculations. *ACS Appl. Mater. Interfaces* **2023**, 12936–12945.

(19) Zhang, T.; Li, L.; Huang, T.; Wan, H.; Chen, W.-Y.; Yang, Z.-X.; Huang, G.-F.; Hu, W.; Huang, W.-Q. Correlation between Spin State and Activity for Hydrogen Evolution of $PtN_2$ Monolayer. *Appl. Phys. Lett.* **2024**, *124* (6), 063903.

(20) Lin, C.; Batchelor-McAuley, C.; Laborda, E.; Compton, R. G. Tafel–Volmer Electrode Reactions: The Influence of Electron-Transfer Kinetics. *J. Phys. Chem. C* **2015**, *119* (39), 22415–22424.

(21) Chen, Z. W.; Li, J.; Ou, P.; Huang, J. E.; Wen, Z.; Chen, L.; Yao, X.; Cai, G.; Yang, C. C.; Singh, C. V.; Jiang, Q. Unusual Sabatier Principle on High Entropy Alloy Catalysts for Hydrogen Evolution Reactions. *Nat. Commun.* **2024**, *15* (1), 359.

(22) Chen, Z.; Liu, Z.; Xu, X. Dynamic Evolution of the Active Center Driven by Hemilabile Coordination in $Cu/CeO_2$ Single-Atom Catalyst. *Nat. Commun.* **2023**, *14* (1), 2512.

(23) Nørskov, J. K.; Bligaard, T.; Logadottir, A.; Kitchin, J. R.; Chen, J. G.; Pandelov, S.; Stimming, U. Trends in the Exchange Current for Hydrogen Evolution. *J. Electrochem. Soc.* **2005**, *152* (3), J23.

(24) Liu, T.; Zhao, X.; Liu, X.; Xiao, W.; Luo, Z.; Wang, W.; Zhang, Y.; Liu, J.-C. Understanding the Hydrogen Evolution Reaction Activity of Doped Single-Atom Catalysts on Two-Dimensional $GaPS_4$ by DFT and Machine Learning. *J. Energy Chem.* **2023**, *81*, 93–100.

(25) Zhang, X.; Chen, J.; Wang, H.; Tang, Y.; Feng, Y. P.; Chen, Y.; Chen, Z. Dynamic Structural Evolution of Single-Atom Catalysts at the Catalyst–Electrolyte Interface: Insights from Electrochemical Coupled Field. *Nano Lett.* **2025**.

(26) Yang, X.; Song, W.; Liao, K.; Wang, X.; Wang, X.; Zhang, J.; Wang, H.; Chen, Y.; Yan, N.; Han, X.; Ding, J.; Hu, W. Cohesive Energy Discrepancy Drives the Fabrication of Multimetallic Atomically Dispersed Materials for Hydrogen Evolution Reaction. *Nat. Commun.* **2024**, *15* (1), 8216.

(27) Ye, S.; Liu, F.; She, F.; Chen, J.; Zhang, D.; Kumatani, A.; Shiku, H.; Wei, L.; Li, H. Hydrogen Binding Energy Is Insufficient for Describing Hydrogen Evolution on Single-Atom Catalysts. *Angew. Chem. Int. Ed.* **2025**, *64* (23), e202425402.

(28) Ni, B.; Shen, P.; Zhang, G.; Zhao, J.; Ding, H.; Ye, Y.; Yue, Z.; Yang, H.; Wei, H.; Jiang, K. Second-Shell N Dopants Regulate Acidic $O_2$ Reduction Pathways on Isolated Pt Sites. *J. Am. Chem. Soc.* **2024**, *146* (16), 11181–11192.

(29) Pang, Y.; Zhang, X.; Li, P.; Xia, G.-J.; Zong, X.; Liu, Y.; Qu, D.; Zheng, K.; An, L.; Wang, X.; Sun, Z. Developing Dual-Atom Catalysts with Tunable Electron Synergistic Effect via Photoinduced Ligand Exchange Strategy. *ACS Catal.* **2025**, *15* (2), 1061–1072.

(30) Koudakan, P. A.; Wei, C.; Mosallanezhad, A.; Liu, B.; Fang, Y.; Hao, X.; Qian, Y.; Wang, G. Constructing Reactive Micro‐Environment in Basal Plane of $MoS_2$ for pH‐Universal Hydrogen Evolution Catalysis. *Small* **2022**, 2107974.

(31) Li, L.; Yang, Z.-X.; Huang, T.; Wan, H.; Nie, J.; Huang, G.-F.; Hu, W.; Huang, W.-Q. Out-of-Plane Polarization Engineering: Optimizing Exciton Dissociation and Solar-to-Hydrogen Efficiency in Photocatalytic Water Splitting. *Appl. Phys. Lett.* **2025**, *126* (18), 183903.

(32) Cui, W.-G.; Gao, F.; Na, G.; Wang, X.; Li, Z.; Yang, Y.; Niu, Z.; Qu, Y.; Wang, D.; Pan, H. Insights into the pH Effect on Hydrogen Electrocatalysis. *Chem. Soc. Rev.* **2024**, *53* (20), 10253–10311.

(33) Frankerl, M.; Patera, L. L.; Giselbrecht, F.; Frederiksen, T.; Repp, J.; Donarini, A. Substrate Polarization Alters the Jahn-Teller Effect in a Single Molecule. *Phys. Rev. Lett.* **2025**, *134* (17), 176203.

(34) Ma, H.; Ye, X.; Li, X.; Xu, Z. J.; Sun, Y. Ferroelectric Polarization Effects of Single-Atom Catalysts on Water Oxidation. *Adv. Mater.* **2025**, *37* (21), 2500285.

(35) Ju, L.; Tan, X.; Mao, X.; Gu, Y.; Smith, S.; Du, A.; Chen, Z.; Chen, C.; Kou, L. Controllable $CO_2$ Electrocatalytic Reduction via Ferroelectric Switching on Single Atom Anchored $In_2Se_3$ Monolayer. *Nat. Commun.* **2021**, *12* (1), 5128.

(36) Yu, G.; Ji, J.; Chen, Y.; Xu, C.; Xiang, H. J. Symmetry Strategy for Rapid Discovery of Abundant Fractional Quantum Ferroelectrics. *Phys. Rev. Lett.* **2025**, *134* (1), 16801.

(37) Li, L.; Yang, Z.-X.; Huang, T.; Wan, H.; Chen, W.-Y.; Zhang, T.; Huang, G.-F.; Hu, W.; Huang, W.-Q. Doping-Free Janus Homojunction Solar Cell with Efficiency Exceeding 23%. *Appl. Phys. Lett.* **2024**, *125* (22), 223904.

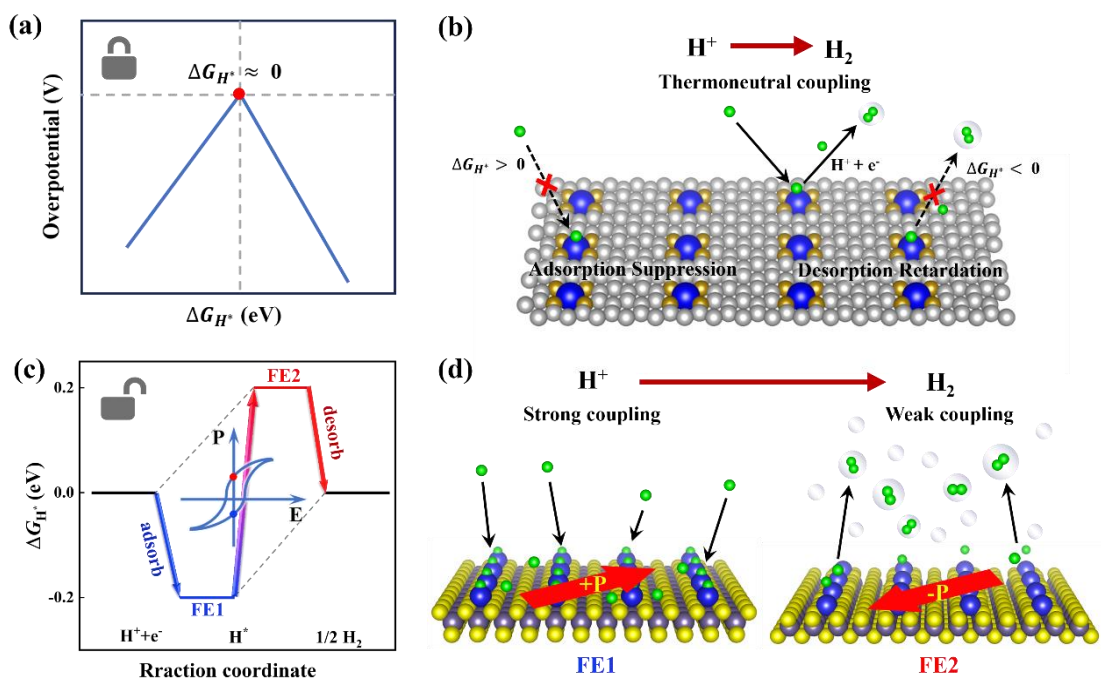

**Figure 1. Decoupling the adsorption/desorption processes to break the Sabatier constraint.** (a) The energy barrier dilemma constrains traditional HER processes. (b) Schematic illustrating the problem of adsorption/desorption suppression in HER catalysts. (c) Principle of free energy oscillation between strong and weak adsorption states driven by polarization potential fields. (d) Schematic illustration of spontaneous hydrogen adsorption/desorption processes induced by ferroelectric domain switching.

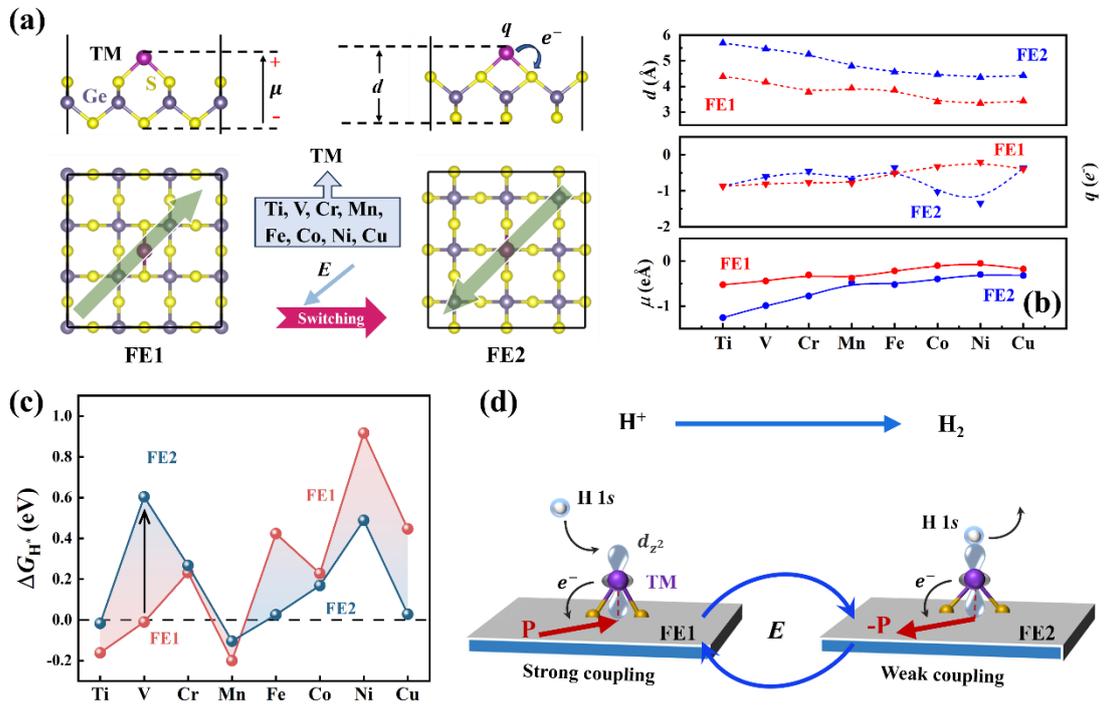

**Figure 2. Ferroelectric modulation of polarization parameters and decoupling of $\Delta G_{H^*}$.** (a) Structures in two polarization states, with light green arrows indicating the in-plane polarization direction. (b) Changes in key material properties under polarization switching, including atomic layer thickness, TM-site Bader charge, and vertical dipole moment. (c) Variation of free energy values under different polarization states. (d) Orbital interactions between active sites and intermediates during ferroelectric modulation.

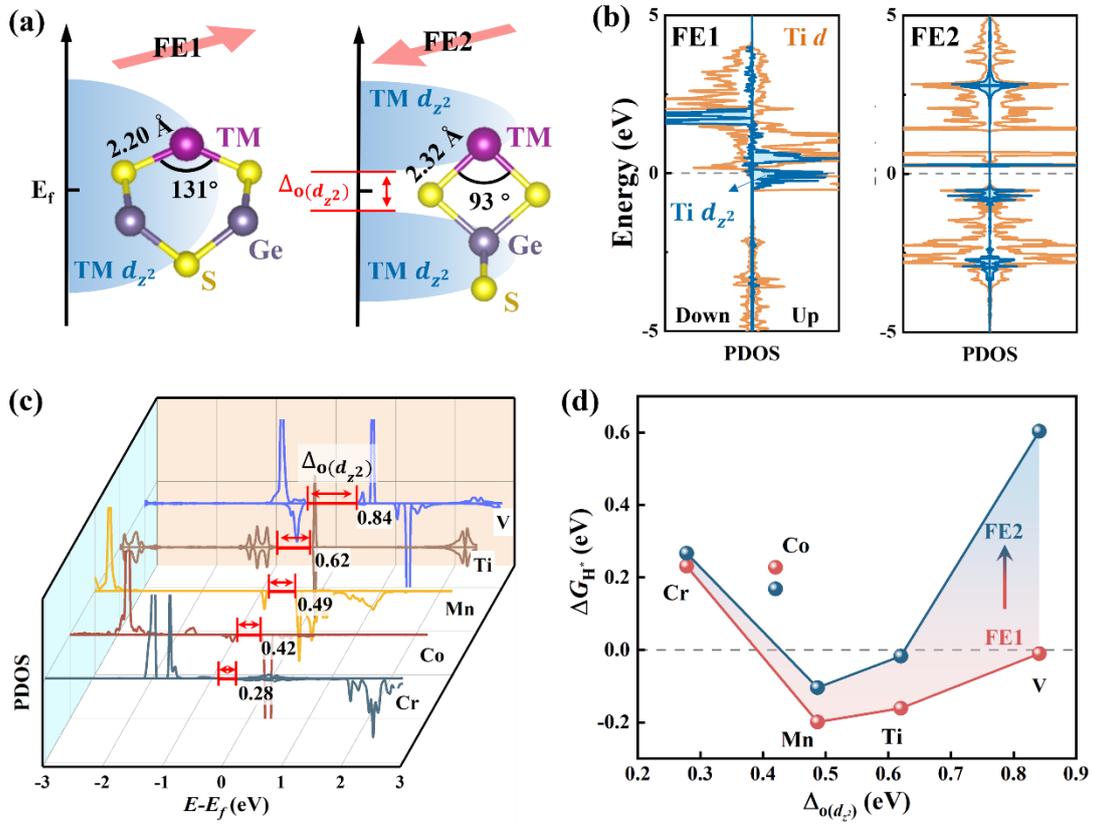

**Figure 3. Polarization-induced $d_{z^2}$ orbital splitting.** (a) The changes in geometric structure and the $d_{z^2}$ orbital splitting ($\Delta_{o(d_{z^2})}$) of TM atom under different polarization. (b) Reconstruction of spin-polarized PDOS in Ti@GeS$_2$. (c) The magnitude of $\Delta_{o(d_{z^2})}$ in different systems. (d) Correlation between the $\Delta_{o(d_{z^2})}$ and $\Delta G_{H^*}$ under different polarization states. Here, the arrows indicate polarization switching, the longer the background color, the greater the change in free energy.

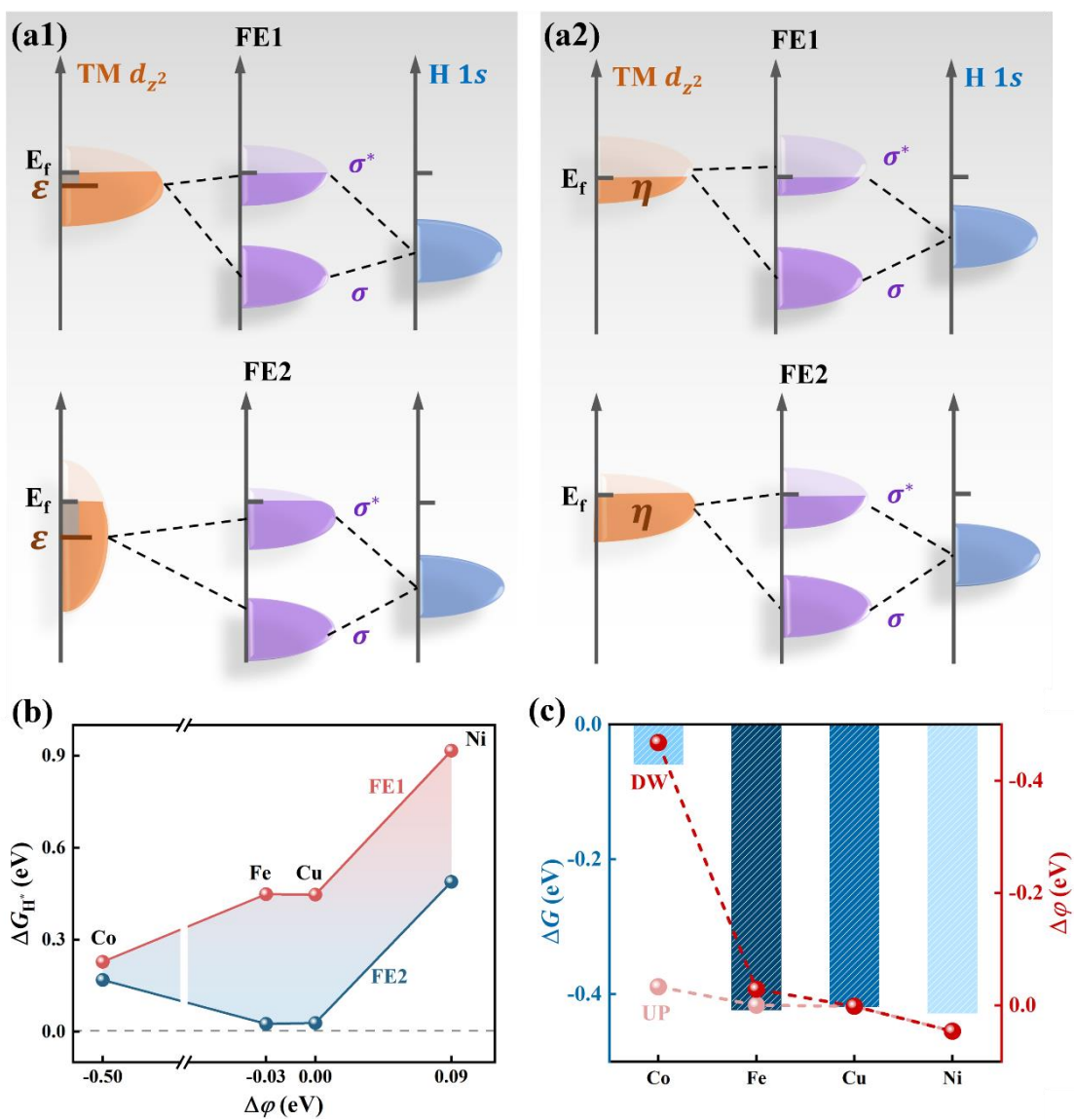

**Figure 4. Synergistic Coupling of $d_{z^2}$-orbital band center ($\varepsilon$) and the occupancy ratio ($\eta$) as an Electronic Structure Descriptor for Electrocatalytic Performance.** Orbital interaction evolution between the TM $d_{z^2}$ and H $1s$ orbital of H versus (a1) $\varepsilon$ and (a2) $\eta$ under distinct polarization states. (b) Correlation between the coupling descriptor $\Delta\varphi$ and the $\Delta G_{H^*}$. (c) Spin-resolved $\Delta\varphi$ versus the free energy difference $\Delta G$, with the spin-down channel (dark red point) exhibiting dominant behavior.

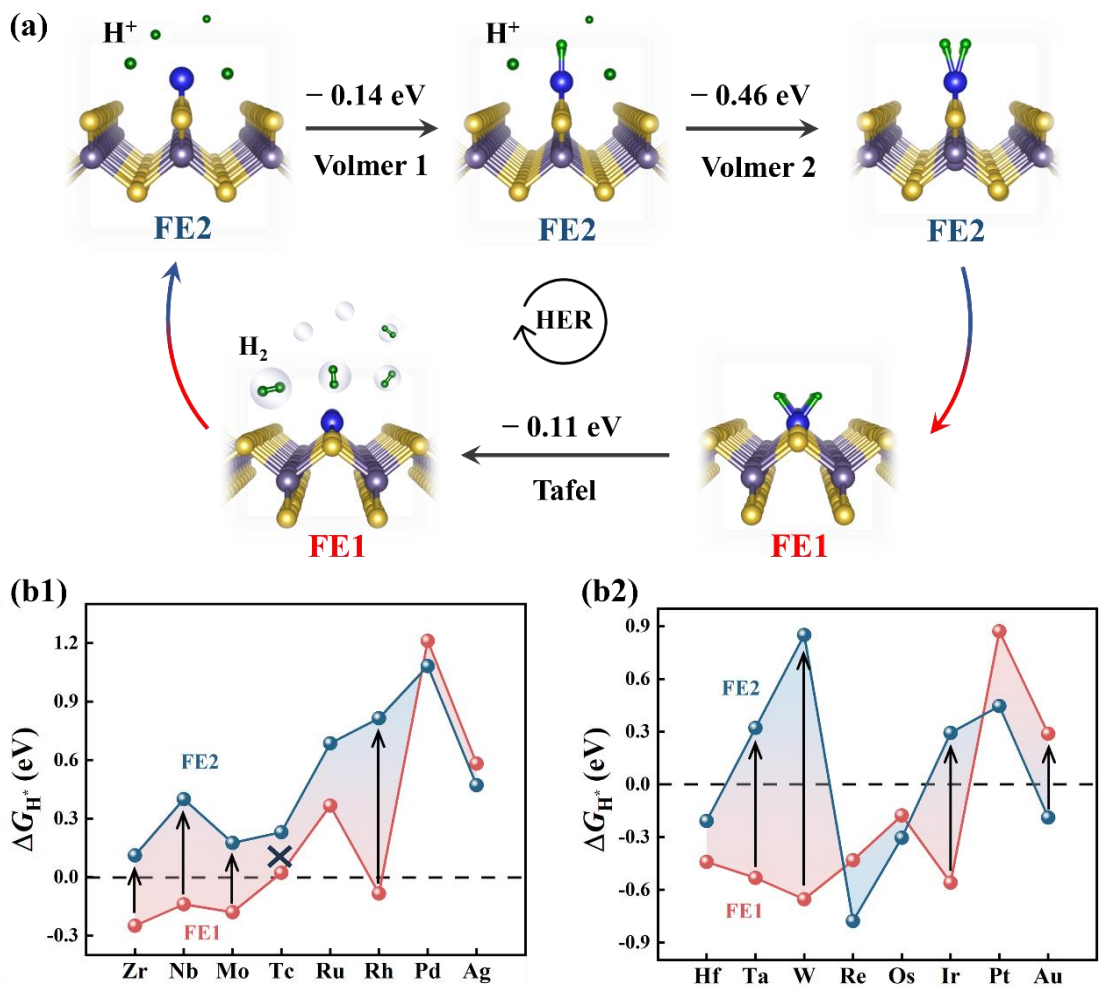

**Figure 5. Decoupling of multiple processes via polarization switching.** (a) The optimal HER pathway in the Fe@GeS$_2$ system; wherein the adsorption process occurs at the FE2 polarization state, and the desorption process takes place at the FE1 polarization state. $\Delta G_{H^*}$ profiles for (b1) Group V elemental atoms and (b2) Group VI elemental atoms. Black arrows indicate the transition of adsorption free energy from negative to positive values.